\documentstyle[pra,aps,epsf,preprint]{revtex}

\newcommand{\bra}[1]{\mathop{\left\langle #1 \right|}\nolimits}
\newcommand{\ket}[1]{\mathop{\left| #1 \right\rangle}\nolimits}

\begin{document}

\tighten 

\title{Analysis of radiatively stable entanglement \\in a system of
two dipole-interacting three-level atoms}
\author{I.\ V.\ Bargatin, B.\ A.\ Grishanin, and V.\ N.\ Zadkov}
\address{International Laser Center and Department of Physics,\\
M.\ V.\ Lomonosov Moscow State University,  Moscow 119899, Russia}
\date{\today}
\maketitle

\begin{abstract}
We explore the possibilities of creating {\em radiatively stable} entangled
states of two three-level dipole-interacting atoms in a $\Lambda$ configuration
by means of laser biharmonic continuous driving or pulses. We propose three
novel schemes for generation of entangled states which involve only the lower
states of the $\Lambda$ system, not vulnerable to radiative decay. Two of them
employ coherent dynamics to achieve entanglement in the system, whereas the
third one uses optical pumping, i.e.,\ an essentially incoherent process.
\end{abstract}

\narrowtext

The concept of quantum entanglement, one of the most intriguing properties of
multipartite quantum systems \cite{bellbook}, has been intensively exploited
over the last decade in connection with quantum information processing. It has
been shown that the use of entangled states opens new horizons in such
practical fields as cryptography \cite{crypt}, computing \cite{steanerev},
information transmission \cite{quantumchannel}, and precision measurement
\cite{spectr}. However, all of these applications become possible only with a
reliable source of entanglement. Traditionally, entangled particles have been
generated in the down-conversion nonlinear process \cite{klyshkobook,bellexp},
but this method is in some cases disadvantageous due to the speedy nature of
the produced particles (photons) and the intrinsic randomness of their
appearance times. That is why efforts are now being made to find ways for
controlled production of entangled states of less volatile {\em massive
particles} \cite{massive}. During the last few years, various methods for
creation of entangled states of {\em atoms}, ranging from continuous
observation of radiative decay \cite{cavlossent,interfent} to controlled cold
collisions \cite{coldcollent}, have been proposed and some of them
experimentally demonstrated \cite{determent,NMRent}.

Though the resonant dipole-dipole interaction (RDDI) was suggested for
realization of entangling dynamics as early as 1995 \cite{barenco95}, it was
only recently that several authors \cite{brennen98,trieste,plenio99}
investigated this interaction in more detail as a method for entangling neutral
atoms in optical traps (neutral atom realizations benefit from the fact that
neutral atoms are less sensitive to stray EM fields---a major source of
decoherence in ions \cite{iontraps}). While the authors of Ref.
\cite{brennen98} offered qualitative arguments for realization of this idea in
dipole traps \cite{grimm99}, the papers \cite{trieste,plenio99} considered
quantitative models of creation of maximally entangled states of two-level
atoms. Unfortunately, such entangled states of two-level atoms have short
lifetimes due to radiative decay. Obviously, since radiative decay and the RDDI
have the same physical nature, we cannot avoid the former while making use of
the latter. In this paper we solve this conceptual problem by presenting new
methods for creation of radiatively stable entanglement in a system of
dipole-interacting {\em three-level} atoms. Though the model considered here is
still far from representing the real situations (see \cite{brennen99} for
details of a possible experimental realization), it offers new insights into
how the RDDI can be used to entangle real, multilevel atoms.

\section{The model}
\label{model}

\begin{figure}
\begin{center}
\epsfxsize=7cm \epsfclipon \leavevmode \epsffile{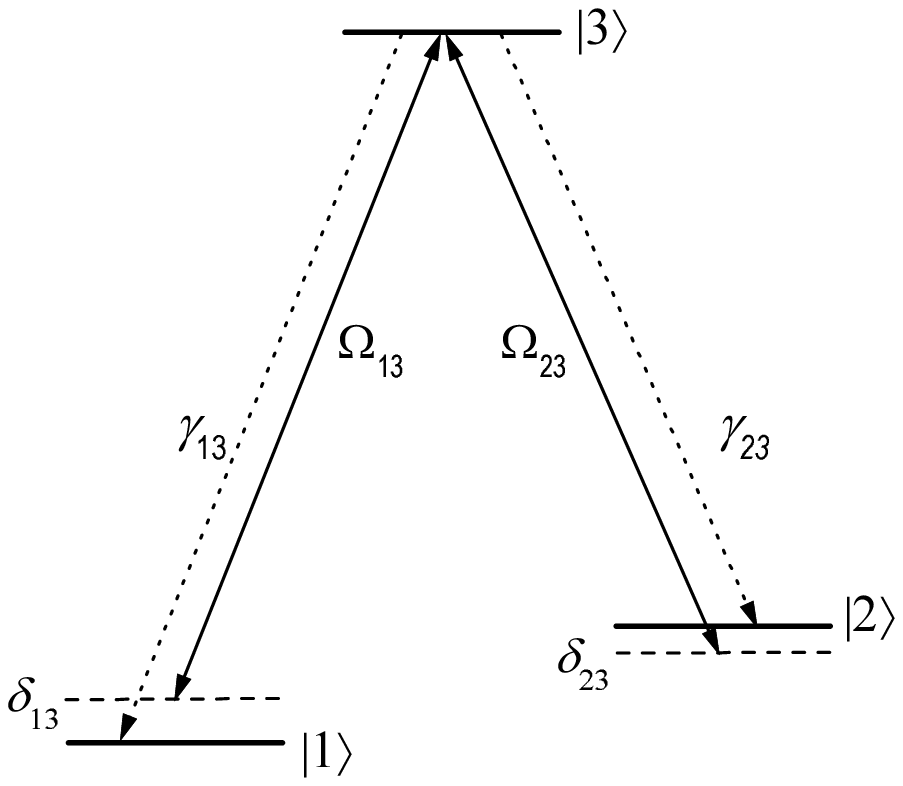}
\end{center}
\caption{The level structure of an isolated three-level atom in a $\Lambda$
configuration. The dipole transitions, $\ket{1} \leftrightarrow \ket{2}$ and
$\ket{2}\leftrightarrow\ket{3}$, are driven by two laser fields, which are
detuned by $\delta_{13}$ and $\delta_{23}$, respectively. Dotted lines show
radiative decay channels and their corresponding rates.} \label{fig:lambda}
\end{figure}

\begin{figure}
\begin{center}
\epsfxsize=8cm \epsfclipon \leavevmode \epsffile{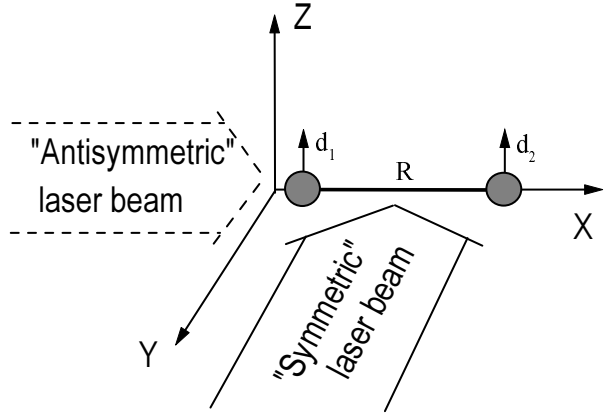}
\end{center}
\caption{Geometry of the model with directions of laser beams for the
``symmetric'' and ``antisymmetric'' laser beams.} \label{fig:geometry}
\end{figure}

Extending the model described in \cite{trieste},  we consider here two
identical {\em three-level} atoms in a $\Lambda$ configuration (Fig.\
\ref{fig:lambda}) fixed at a distance $R$. The dipole transitions $\ket1
\leftrightarrow \ket3$ and $\ket2 \leftrightarrow \ket3$ of both atoms are
driven by two near-resonant laser fields. Taking the two limiting cases, we
consider only two types of geometry: when the laser fields are either
perpendicular or parallel to the radius vector $\vec R$ connecting the atoms
(these geometries are shown in Fig.\ \ref{fig:geometry} and identified as
symmetric and antisymmetric, respectively). Within the interaction picture and
rotating wave approximation, the evolution of the system interacting with the
laser fields is governed by the following master equation \cite{kurizki87}:
\begin{equation}\label{ME}
  \frac{\partial \hat \rho}{\partial t}=
  -\frac{i}{\hbar}\left[\hat{\cal H}_{\rm eff},\hat\rho\right]+
  \sum_{i,j,k=1,2}\frac{\gamma_{k3}^{(ij)}}{2}\left(2\hat \sigma_{3k}^{(i)}
  \hat\rho\hat \sigma_{k3}^{(j)} -
  \hat\rho\hat \sigma_{3k}^{(i)}\hat
  \sigma_{k3}^{(j)} - \hat \sigma_{3k}^{(i)}\hat \sigma_{k3}^{(j)}\hat\rho
  \right),
\end{equation}

\noindent where the upper indices, $i$ and $j$, number the atoms, the lower
ones, $k3$ and $3k$ ($k=1,\,2$), refer to dipole transitions of the atoms, and
$\sigma_{kl}^{(i)}$ denotes the Heisenberg transition operators from level
$\ket{k}$ to level $\ket{l}$ within the $i$th atom. Relaxation effects in the
system are characterized by the single-atom decay rates, $\gamma_{k3} =
\gamma_{k3}^{(11)} = \gamma_{k3}^{(22)}$, which correspond to the conventional
radiative decay into free space, and the photon exchange rates,
$\gamma_{k3}^{(12)}= \gamma_{k3}^{(21)}$, which describe collective relaxation,
a well-known companion of the RDDI. The effective Hamiltonian $\hat{\cal
H}_{\rm eff}$ includes interaction with the laser field and the RDDI coupling
on both transitions:
\begin{equation}\label{Heff3L}
\hat{\cal H}_{\rm eff}=\hbar\sum_{i,k=1,2}\left(\delta_{k3} \hat n_{k}^{(i)}+
\frac{\Omega_{k3}^{(i)}}{2} \hat \sigma_{k3}^{(i)}+ \chi_{k3}\hat
\sigma_{k3}^{(1)}\hat \sigma_{3k}^{(2)}+ \mbox{H.\,c.}\right),
\end{equation}

\noindent where $\hat n_{k}^{(i)}$ stands for the population operator of the
level $\ket{k}$ in the $i$th atom, $\delta_{k3}$ are the detunings of the laser
field frequencies from the corresponding transitions
$\ket{k}\leftrightarrow\ket{3}$ of an isolated atom, $\Omega_{k3}^{(i)}$ is the
Rabi frequency of the laser field acting on the $\ket{k}\leftrightarrow\ket{3}$
transition of the $i$th atom, and $\chi_{k3}$ is the RDDI coupling strength on
the $\ket{k}\leftrightarrow\ket{3}$ transition. Throughout the rest of this
paper we will consider the case of wide homogeneous laser beams, so that the
Rabi frequencies acting on the two atoms may differ in phase but not in
magnitude, $|\Omega_{k3}^{(1)}| =|\Omega_{k3}^{(2)}|$.

Normalizing the RDDI parameters, $\chi_{k3}$, $\gamma_{k3}^{(12)}$, and
$\gamma_{k3}^{(21)}$, by the decay rate of an isolated atom, $\gamma_{k3}$, we
introduce the dimensionless parameters
\begin{equation}\label{dimpar}
g_{k3}=\gamma_{k3}^{(12)}/\gamma_{k3}=\gamma_{k3}^{(21)}/\gamma_{k3},
\quad f_{k3}=\chi_{k3}/\gamma_{k3},
\end{equation}

\noindent which are given by the following expressions
\cite{kurizki87}:
\begin{equation}\label{f&g}
\renewcommand{\arraystretch}{1.5}
\begin{array}{ll}
 f_{k3}=F(\varphi_{k3})=& \displaystyle
\frac{3}{2}\left(\frac{\cos\varphi_{k3}}{\varphi_{k3}^3}+\frac{\sin
  \varphi_{k3}}{\varphi_{k3}^2}-\frac{\cos \varphi_{k3}}{\varphi_{k3}}\right)
  \left[\vec e_1 \vec e_2 -(\vec e_1 \vec e_R)(\vec e_2 \vec
  e_R)\right]\\
  & \displaystyle -3\left(\frac{\cos \varphi_{k3}}{\varphi_{k3}^3}+\frac{\sin
  \varphi_{k3}}{\varphi_{k3}^2}\right)\left[(\vec e_1 \vec e_R)
  (\vec e_2 \vec e_R)\right],\\
  g_{k3}=G(\varphi_{k3}) =&\displaystyle \frac{3}{2}\left(\frac{\sin
\varphi_{k3}}{\varphi_{k3}}+\frac{\cos
  \varphi_{k3}}{\varphi_{k3}^2}-\frac{\sin \varphi_{k3}}{\varphi_{k3}^3}\right)
  \left[\vec e_1 \vec e_2 -(\vec e_1 \vec e_R)(\vec e_2 \vec
  e_R)\right]\\
   &\displaystyle +3\left(\frac{\sin \varphi_{k3}}{\varphi_{k3}^3}-
  \frac{\cos\varphi_{k3}}{\varphi_{k3}^2}\right)\left[(\vec e_1 \vec e_R)
  (\vec e_2 \vec e_R)\right],
\end{array}
\end{equation}

\noindent where $\vec e_i$ ($i=1,2$) is the unit vector in the direction of the
dipole moment matrix element of the corresponding transition
$\ket{k}\leftrightarrow \ket{3}$ of the $i$th atom, $\vec e_R$ is the unit
vector in the direction of $\vec R$, and $\varphi_{k3}=k_{k3} R$ is the
dimensionless distance between the atoms ($k_{k3}=\omega_{k3}/c$ is the wave
number associated with the transition $\ket{k}\leftrightarrow \ket{3}$ of an
isolated atom). Throughout the following discussion we will assume, for the
sake of simplicity, that the dipole moments are real, collinear with each
other, and perpendicular to the radius vector $\vec R$ (other dipole moment
orientations lead to qualitatively the same results).

\begin{figure}
\begin{center}
\epsfxsize=\textwidth \epsfclipon \leavevmode \epsffile{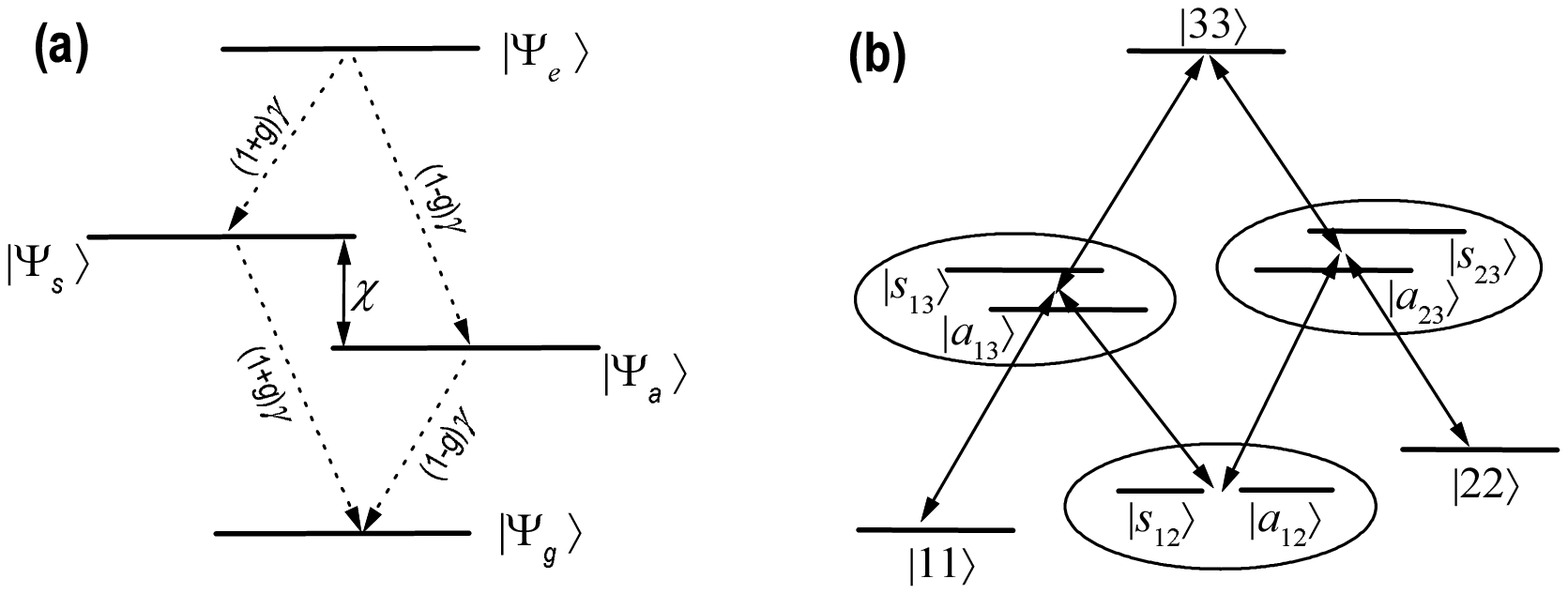}
\end{center}
\caption{(a) Energy levels of two dipole-interacting two-level atoms (Dicke
states). The two maximally entangled Dicke states, $\ket{\Psi_a}$ and
$\ket{\Psi_s}$, are split by the RDDI coupling strength $\chi$. Also shown in
the figure are the radiative decay channels with the corresponding decay rates.
(b) The same for two dipole-interacting $\Lambda$ systems. Shown are the
doublets, formed by symmetric and antisymmetric Dicke-like states, and the
laser-induced transitions induced by the two components of the biharmonic
driving.} \label{fig:levels}
\end{figure}

In the case of two-level atoms, the simplest description of the system dynamics
is offered by the basis of the Dicke states, which is formed by the doubly
excited state, $\ket{\Psi_e}= \ket{e}_1\ket{e}_2$, the ground state,
$\ket{\Psi_g}= \ket{g}_1\ket{g}_2$, and the two singly excited maximally
entangled states---the symmetric, $\ket{\Psi_s}=
\frac{1}{\sqrt{2}}(\ket{g}_1\ket{e}_2+\ket{e}_1\ket{g}_2)$, and the
antisymmetric one, $\ket{\Psi_a}=\frac{1}{\sqrt{2}} (\ket{g}_1 \ket{e}_2-
\ket{e}_1\ket{g}_2)$ [the corresponding energy diagram is shown in
Fig.~\ref{fig:levels}(a)]. For the case of three-level atoms considered here it
is useful to introduce simple generalizations of the Dicke states. The role of
the ground and doubly excited Dicke states is then played by the three tensor
product states $\ket{kk}= \ket{k}_1 \ket{k}_2,$ $k=1,2,3,$ while the symmetric
and antisymmetric Dicke states now are represented by the three symmetric and
three antisymmetric maximally entangled states $\ket{s_{kl}}=
\frac{1}{\sqrt{2}}(\ket{k}_1\ket{l}_2+\ket{k}_1\ket{l}_2)$ and
$\ket{a_{kl}}=\frac{1}{\sqrt{2}}(\ket{k}_1\ket{l}_2-\ket{k}_1\ket{l}_2),$
$k,l=1,2,3;$ $k<l$. The corresponding energy diagram is shown in
Fig.~\ref{fig:levels}(b). Note that in both two- and three-level models the
energy levels can be grouped according to their type of symmetry: the
unentangled states $\ket{kk}=\ket{k}_1\ket{k}_2,$ as well as the states
$\ket{s_{kl}}=\frac{1}{\sqrt{2}} (\ket{k}_1 \ket{l}_2+\ket{k}_1\ket{l}_2)$, can
be said to belong to one type of symmetry (symmetric with respect to the atom
interchange), and the states $\ket{a_{kl}}=\frac{1}{\sqrt{2}}
(\ket{k}_1\ket{l}_2- \ket{k}_1 \ket{l}_2)$ to another (antisymmetric with
respect to the atom interchange). The transitions between these levels can then
be classified as symmetry-preserving and symmetry-breaking, respectively. It is
easy to show that, due to the form of the transition matrix elements, the
symmetry-preserving transitions are sensitive only to the sum of the Rabi
frequencies, $\Omega_{k3}^{(1)}+\Omega_{k3}^{(2)}$, acting on the atoms, and
the symmetry-breaking transitions only to their difference,
$\Omega_{k3}^{(1)}-\Omega_{k3}^{(2)}$.

In the following, we also assume that the system is initially stored in the
$\ket{11}$ state, which can easily be achieved by conventional optical pumping
methods \cite{optpumpref}.

\section{Coherent entangling processes}

\subsection{Resonant Raman pulses}

In our previous paper \cite{trieste} we have shown that the maximally entangled
Dicke states $\ket{\Psi_s}$ or $\ket{\Psi_a}$ of two two-level atoms can be
efficiently populated at small interatomic distances simply by applying an
appropriately tailored laser pulse. Assuming that initially the entire
population of the system is concentrated in the ground state $\ket{\Psi_g}$,
this pulse should be tuned into resonance with a transition to only one of
these maximally entangled states. Then, by applying a $pi$ pulse analog, a
significant part of the population of the system can be transferred to one of
these states, thereby creating entanglement in the system. We have also shown
that the entanglement fidelity, defined as the population of the corresponding
maximally entangled state, can be made arbitrarily close to unity as the
interatomic distance $R$ goes to zero.

In this paper we propose new ways to create {\em stable} entanglement in a
system of two three-level atoms. To be radiatively stable, the created
entangled states should involve only the lower levels $\ket{1}$ and $\ket{2}$
of the original $\Lambda$ system of each atom, as only these states are not
vulnerable to radiative decay. Therefore, our goal here will be to achieve the
maximum possible population of one of the maximally entangled states
$\ket{a_{12}}$ or $\ket{s_{12}}$ (see Fig.\ \ref{fig:levels}(b)). The most
straightforward way to do this is to extend the results of the two-level model
to the three-level one, considered here, by using resonant Raman pulses. By the
latter we mean a sequence of two coherent $pi$ pulses, the first of which
transfers the population to one of the maximally entangled states involving the
initial lower level of the $\Lambda$ system and some quickly decaying upper
lying ``transit'' level, while the second one transfers the entire population
of the ``transit'' level to another radiatively stable lower level of the
$\Lambda$ system, thus removing the radiative instability of the entanglement.

In the considered system of two dipole-interacting three-level atoms, the role
of the intermediate ``transit'' state can be played by the above-mentioned
levels $\ket{a_{13}}$ or $\ket{s_{13}}$ (one should not forget that the system
is initially in the state $\ket{11}$). During the first step, the pulse
resonant, for example, with the $\ket{11}\leftrightarrow \ket{s_{13}}$
transition transfers the population to the maximally entangled state
$\ket{s_{13}}$; the second step creates the radiatively stable maximally
entangled state $\ket{s_{12}}$ by application of the symmetry-preserving $pi$
pulse resonant with the $\ket{s_{13}}\leftrightarrow\ket{s_{12}}$ transition.
In fact, it is the symmetry preservation rules that prevent population from
going into the $\ket{a_{13}}$ state in a transition that is also resonant with
the second pulse.

For both pulses to be resonant, the parameters of the laser field
should be chosen in the following way
\begin{equation}\label{rampar1}
\alpha_{k3}=0, \quad \delta_{k3}=\chi_{13}/2, \quad
|\Omega^{(i)}_{k3}|\ll|\chi_{13}|,
\end{equation}

\noindent where $\alpha_{k3}$ is the phase difference between the Rabi
frequencies acting on the two atoms, $\Omega_{k3}^{(1)}= \Omega_{k3}^{(2)}
\exp(i\alpha_{k3})$ (considering laser beams formed by traveling waves,
$\alpha_{k3}$ varies from zero for the symmetric geometry to $\varphi_{k3}$ for
the antisymmetric one, and takes all the intermediate values for other types of
laser field geometry). The parameters given by (\ref{rampar1}) correspond
therefore to the case when both lasers are used in the symmetric geometry, the
``transit'' state is $\ket{s_{13}}$, and the final radiatively stable maximally
entangled state is $\ket{s_{12}}$. Other types of geometries and laser
parameter sets can obviously be chosen when using the other intermediate state
$\ket{a_{13}}$ and/or creating the other radiatively stable maximally entangled
state $\ket{a_{12}}$. For example, to create the $\ket{a_{12}}$ state, one can
use the following set of parameters:
\begin{equation}\label{rampar2}
\alpha_{13}=\varphi_{13}, \quad \alpha_{23}=0, \quad
\delta_{k3}=-\chi_{13}/2, \quad |\Omega^{(i)}_{k3}|\ll|\chi_{13}|.
\end{equation}

The phase differences $\alpha_{k3}$ in this case correspond to one of the
lasers beams being used in the antisymmetric geometry, and the other in the
symmetric one. Note that, when using antisymmetric geometry at small
interatomic distances ($\varphi_{k3}\ll 1$), most of the laser power is
``wasted'' since only a fraction of it contributes to the corresponding
transition matrix element $\left|\bra{11}\hat {\cal H}_{\rm
eff}/\hbar\ket{a_{13}}\right| =|\Omega_{13}^{(1)}-\Omega_{13}^{(2)}|/2=
|\Omega_{13}^{(i)}|\sin(\varphi_{13}/2)\ll |\Omega^{(i)}_{13}|$, and actually
induces transitions $\ket{11}\leftrightarrow\ket{a_{13}}$.

While a simple estimate of the resulting fidelity of creation of the maximally
entangled state is offered by a product of the fidelities of each step of the
resonant Raman process (which were calculated in \cite{trieste} within the
two-level atoms model), rigorous results can be obtained only by explicit
solution of the corresponding master equation. Due to the high dimensionality
of the master equation (\ref{ME}), this calculation is rather demanding
computationally, and was not included in the present treatment.

\subsection{Stimulated Raman adiabatic passage}

Another coherent method for creation of maximally entangled states is based on
the stimulated Raman adiabatic passage (STIRAP) technique \cite{bergmann98}, a
well-known alternative to Raman pulses. The STIRAP method uses adiabatic
following of the system state after the slowly changing parameters of the laser
field, which are chosen to form a so-called counterintuitive pulse sequence.
The STIRAP technique benefits from extremely low probabilities of losing
coherence due to radiative decay of the intermediate states, and has already
been proposed for use in entanglement-related problems \cite{pellizzari95}. In
our case, efficient transfer of the population from the state $\ket{11}$ to the
state $\ket{a_{12}}$ or $\ket{s_{12}}$ may be deterred by existence of several
intermediate states \cite{vitanov99}. However, as we show below, efficient
transfer is still possible for appropriately chosen laser pulse parameters.

To realize STIRAP in our system we need to choose the frequencies and
geometries of the two constituent laser pulses in a way that would leave active
(i.e.,\ resonant and having strong transition amplitudes due to the use of the
corresponding geometries, see section \ref{model}) only two transitions in the
whole system. An appropriate choice is given by
\begin{equation}\label{stirappar}
\alpha_{13}=0, \quad\alpha_{23}=\pi, \quad \delta_{k3}=\chi_{13}/2, \quad
\max_t |\Omega_{k3}^{(i)}|=\Omega_0\ll|\chi_{13}|,
\end{equation}

\noindent where $\Omega_0$ stands for the amplitude of the corresponding
constituents of the counterintuitive laser pulse sequence. The condition
$\alpha_{23}=\pi$, which is very important as it prevents leakage of population
into other levels, can be easily realized by using two laser beams in
antisymmetric geometry, which form a standing wave with one of the nodes
situated exactly in the middle of the vector $\vec R$ connecting the two atoms
\cite{plenio99}. In this case only two transitions,
$\ket{11}\leftrightarrow\ket{s_{13}}$ and
$\ket{s_{13}}\leftrightarrow\ket{a_{13}}$, are active, and the adiabatic
passage results in transfer of the total population to the radiatively stable
state $\ket{a_{13}}$.

\begin{figure}
\begin{center}
\epsfxsize=\textwidth \epsfclipon \leavevmode \epsffile{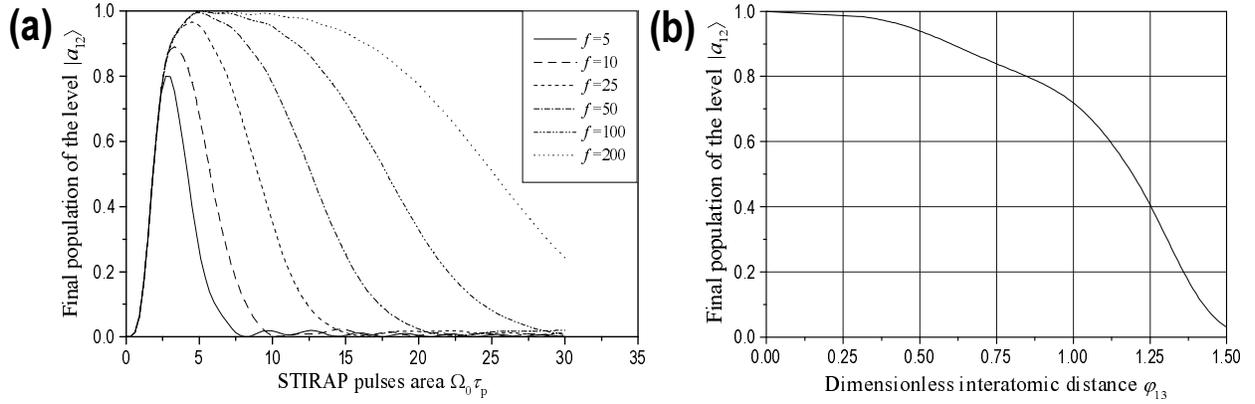}
\end{center}
\caption{(a) Population of the maximally entangled state $\ket{a_{12}}$ after
adiabatic passage versus the pulse area for different values of the RDDI
parameter $f=f_{13}=\chi_{13}/\gamma_{13}$. (b) The same for the optimal value
of the laser pulses area, $\Omega_0\tau_p$, versus the interatomic distance
$\varphi_{13}$. In both graphs we assume equal decay rates of the two channels
of the original $\Lambda$ system, $\gamma_{13}=\gamma_{23}$.}
\label{fig:stirapfid}
\end{figure}

We have numerically calculated the final population (fidelity) of the state
$\ket{a_{12}}$ after the STIRAP procedure by explicit solution of the
corresponding Schr\"odinger equation with the Hamiltonian given by
(\ref{Heff3L}). The two laser field pulses had the same Gaussian form and were
delayed with respect to each other by their length \cite{bergmann98}, and the
rest of the parameters were given by (\ref{stirappar}). For determinacy, the
length of the pulses was chosen to be equal to one-tenth of the lifetime of the
excited level $\ket{3}$ of the original $\Lambda$ system, $\tau_p=
0.1/(\gamma_{13}+ \gamma_{23})$. The final population of the level
$\ket{a_{12}}$ is shown in Fig.~\ref{fig:stirapfid}(a) as a function of the
pulses Rabi frequency amplitude, $\Omega_0$, for different values of the RDDI
splitting parameter $f_{13}$. As one can see from the figure, for sufficiently
high RDDI splittings (i.e.,\ for sufficiently small atomic separations) the
fidelity first grows with increasing $\Omega_0$, reaching saturation at
$\Omega_0\tau_p\approx 5$, which corresponds to the adiabaticity condition on
the pulse area \cite{bergmann98}. Then, after some point, the final inequality
in (\ref{stirappar}) is no longer fulfilled and the efficiency of the process
degrades due to nonresonant excitation of other levels caused by power
broadening. For the same reasons, the fidelity does not reach unity at any
values of $\Omega_0$ for low RDDI splittings (large interatomic separations).
In Fig.~\ref{fig:stirapfid}(b) we show the overall fidelity of the STIRAP
method for optimized values of the Rabi frequency amplitude as a function of
the interatomic distance $\varphi_{13}$.

As we ignore relaxation processes in this model (a common practice for STIRAP
simulations), one should beware of relaxation-induced errors. However, these
errors assume significant values only for the case of long pulses, $\tau_p\geq
1/(\gamma_{13}+\gamma_{23})$, and low overall STIRAP process fidelity, i.e.,\
situations that are not of great concern to us here.

\section{An incoherent entangling process:
Optical pumping}

An interesting alternative to the coherent methods can be offered by optical
pumping schemes where the stationary state of the system corresponds to one of
the maximally entangled states. In this situation, the population of the system
is pumped into the entangled state after asymptotically large time periods.

Consider the following choice of the laser field parameters:
\begin{equation}\label{pumppar1}
\alpha_{k3}=0, \quad \delta_{k3}=\chi_{k3}/2, \quad
|\Omega_{k3}|\ll|\chi_{k3}|.
\end{equation}

\begin{figure}
\begin{center}
\epsfxsize=\textwidth \epsfclipon \leavevmode \epsffile{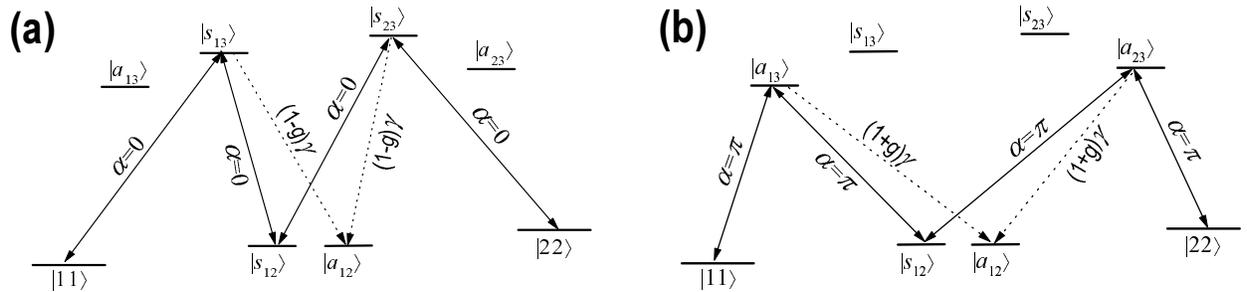}
\end{center}
\caption{Energy level diagram and transitions for two variants, symmetric (a)
and antisymmetric (b), of the optical pumping scheme. Solid lines indicate
laser-induced transitions with the corresponding phase differences $\alpha$
between two atoms. Dotted lines show the significant decay channels with the
corresponding decay rates. The negligibly populated upper state $\ket{33}$ is
omitted in both figures.} \label{fig:optpump}
\end{figure}

\noindent Neglecting nonresonant excitation at small interatomic distances,
only a few transitions remain resonant and have the corresponding geometry.
These active transitions are shown in Fig.~\ref{fig:optpump}(a) (the upper
state $\ket{33}$ is omitted in the figure, as it is only negligibly excited at
small interatomic distances \cite{trieste}).  As seen from the figure, the
maximally entangled state $\ket{a_{12}}$ is not included in the chain formed by
the laser-induced transitions; however, it is still populated as a result of
the decay of the upper-lying levels, as shown by the dotted lines in the same
figure. As the state $\ket{a_{12}}$ is stable with respect to both the
laser-induced transitions and radiative decay, all of the population will
eventually be pumped into this state. One should note, though, that as the
interatomic distance goes to zero, the symmetry-breaking decay rates decrease,
which leads to a corresponding increase of the required pumping time. If we
choose another configuration that uses antisymmetric standing-wave geometry of
the laser beams [Fig.~\ref{fig:optpump}(b)],
\begin{equation}\label{pumppar2}
\alpha_{k3}=\pi, \quad \delta_{k3}=-\chi_{k3}/2, \quad
|\Omega_{k3}|\ll|\chi_{k3}|,
\end{equation}

\noindent the increase of the pumping time is still brought on by the decrease
of efficiency of the symmetry-breaking laser-induced transitions at small
distances since the corresponding transfer matrix elements are proportional to
$|\Omega_{k3}^{(1)}-\Omega_{k3}^{(2)}|\sim \sin(\varphi_{k3}/2)$.

Strictly speaking, the above arguments hold only in the case when the RDDI
coupling constants $\chi_{k3}$ on different transitions are equal (possibly, up
to an error on the order of $\gamma_{k3}$). This condition is satisfied, for
example, when the two lower levels of the original $\Lambda$ system are
sublevels of the same atomic level. However, even when the RDDI coupling on two
transitions differs considerably, the present treatment is still applicable
provided that one uses four lasers instead of two to satisfy all of the
resonance conditions for transitions shown in Fig.\ \ref{fig:optpump}. In
contrast to the methods presented in the previous sections, it is also very
important to {\em avoid} high a degree of mutual coherence of the components of
the biharmonic laser pumping, as otherwise the population of each atom will be
trapped in a corresponding dark state \cite{arimondorev}.

\begin{figure}
\begin{center}
\epsfxsize=\textwidth \epsfclipon \leavevmode \epsffile{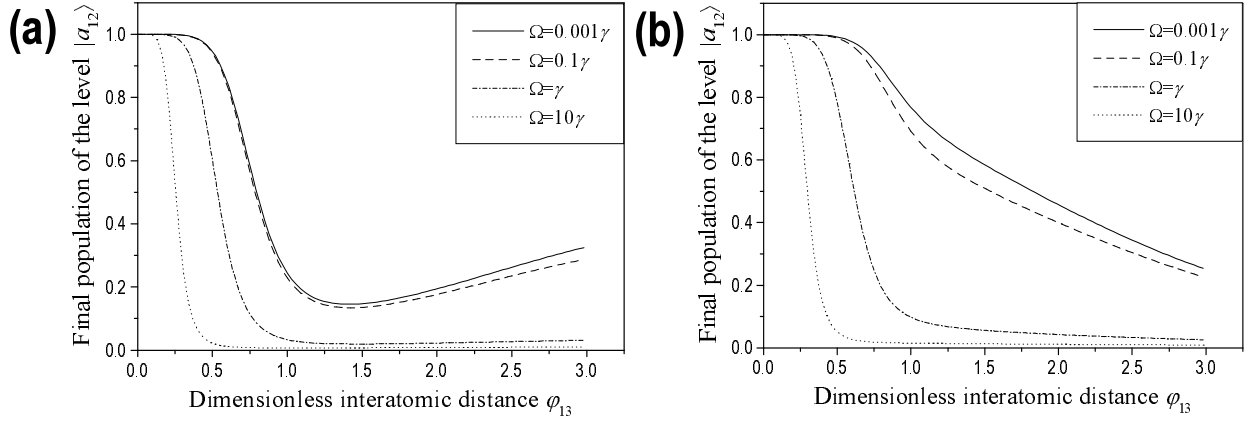}
\end{center}
\caption{Population of the maximally entangled state $\ket{a_{12}}$ in the
stationary solution of the master equation as a function of the interatomic
distance $\varphi_{13}$, for different values of the Rabi frequencies
$|\Omega^{(i)}_{k3}|=\Omega$, $i,k=1,2$ and the two discussed geometries,
symmetric (a) and antisymmetric (b).} \label{fig:pumpfid}
\end{figure}

To prove the foregoing arguments, we have numerically calculated the stationary
states of the master equation (\ref{ME}) with the laser pumping parameters
given by (\ref{pumppar1}) and (\ref{pumppar2}). In order to disrupt trapping of
the populations of the two atoms in the single-atom dark states, we introduced
additional elastic dephasing of the lower level transition
$\ket{1}\leftrightarrow\ket{2}$ in both atoms, which can be easily realized by
the relative jitter of the two pumping laser frequencies. Assuming that this
elastic dephasing is characterized by the rate $\Gamma_{12}$, the corresponding
relaxation superoperator, which should be plugged into the master equation
(\ref{ME}), has the form
\begin{equation}\label{jittersup}
{\cal L}_{\rm jitter}\, \hat \rho=
\Gamma_{12}\sum_{i,j=1,2}\left(2\sigma_z^{(i)}\hat\rho\sigma_z^{(j)}-
\hat\rho\hat\sigma_z^{(i)}\hat\sigma_z^{(j)}-
\hat\sigma_z^{(i)}\hat\sigma_z^{(j)}\hat\rho \right),
\end{equation}

\noindent where $\sigma_z^{(i)}=n_{1}^{(i)}-n_{22}^{(i)}$ is the lower level
population difference operator in the $i$th atom. For simulations we used a
realistic value $\Gamma_{12}=0.01\gamma$, where we again assume for simplicity
$\gamma=\gamma_{13}=\gamma_{23}$. The results of the numerical calculations of
the steady state population of the level $\ket{a_{12}}$ for different values of
laser pumping Rabi frequencies $\Omega$ (in our calculations they are equal for
all transitions and atoms, $|\Omega^{(i)}_{k3}|=\Omega$, $i,k=1,2$) are
presented in Fig.~\ref{fig:pumpfid} as a function of the interatomic distance
$\varphi_{13}$ for the two geometries discussed. As seen from the graphs, the
fidelity of the entanglement produced first monotonically decreases with
increasing Rabi frequency, and then strongly degrades when the magnitude of the
Rabi frequency approaches that of the RDDI splitting due to
power-broadening-induced nonresonant excitation. The graphs for
$\Omega=0.001\gamma$ in Fig.~\ref{fig:pumpfid}, therefore, decently represent
the overall fidelity of the optical pumping method. For low Rabi frequency
amplitudes, the antisymmetric geometry clearly shows better results and
achieves fidelity of 0.8 at $\varphi_{13}\approx 1$, which is much better than
the fidelities achieved by other methods at such distances.

\section{Conclusions}

We have considered three methods for creation of radiatively stable
entanglement in a system of two dipole-interacting three-level atoms in a
$\Lambda$ configuration. It was shown that the radiatively stable maximally
entangled states, $\ket{a_{12}}$ and $\ket{s_{12}}$, which involve only the
lower levels of the original $\Lambda$ systems, can be efficiently populated at
small interatomic distances by employing coherent or incoherent methods.

The first of the coherent methods, which employs resonant Raman pulses for
transfer of population (first to a radiatively unstable maximally entangled
state and then to a stable maximally entangled state), makes use of specific
resonance conditions and symmetry-preservation rules. The second coherent
method, which utilizes a STIRAP process, realizes adiabatic transfer of the
population of the system into the final state coinciding with one of the
radiatively stable maximally entangled states. The STIRAP method, however,
requires the use of standing waves to avoid leakage of population into
unentangled states.

As a rather surprising result, we have also shown that entanglement can be
deterministically created as a result of an incoherent process \cite{ground},
optical pumping in our case. Creating a laser field configuration where one of
the maximally entangled states, $\ket{a_{12}}$ or $\ket{s_{12}}$, is not
included in the chain of laser-induced transitions, we achieve high populations
of that state at asymptotically large times due to radiative decay into that
state. An important restriction for realization of the optical pumping method
is that one has to avoid high mutual coherence of the pumping laser beams, but
this restriction becomes an advantage when realizing the proposed schemes
experimentally as it is usually easier to provide an incoherent pumping than a
coherent one.

For two of the proposed methods (STIRAP and optical pumping), the fidelity of
the created approximations of the maximally entangled states was calculated,
and it shows qualitatively the same dependence on the interatomic distance $R$
as in the previously considered two-level atom model \cite{trieste}. The
fidelity of 0.8 (a good benchmark for Bell inequality violations) is achieved
in all of the considered methods at interatomic separations between
one-fifteenth and one-sixth of the wavelengths of the working transitions.

In conclusion, we have shown that radiatively stable maximally entangled states
can be created in a system of two dipole-interacting atoms under conditions
that can be experimentally implemented, for example, in optical lattices
\cite{brennen98,brennen99}. The general form of the RDDI operator also suggests
that simple analogs of the proposed methods can be employed in other physical
systems, such as quantum dots in semiconductors \cite{ground,qdots2} or cavity
QED systems \cite{pellizzari95,kurizki96} (or, indeed, a combination of the
latter two \cite{divincenzo99}).

\acknowledgements

This work was partially supported by the programs ``Fundamental Metrology" and
``Physics of Quantum and Wave Processes" of the Russian Ministry of Science and
Technology.

\end{document}